\documentclass[12pt]{article}

\usepackage{amssymb,graphicx}

\begin{document}

\begin{center}
  \bf \Large An enigmatic long-lasting $\gamma$-ray burst not accompanied by a
  bright supernova
\end{center}

\bigskip

\noindent M. Della Valle$^{1}$, G. Chincarini$^{2,3}$, N. Panagia$^{4,5,6}$, G.
Tagliaferri$^{3}$, D. Malesani$^{7}$, V. Testa$^{8}$, D. Fugazza$^{3,2}$, S.
Campana$^{3}$, S. Covino$^{3}$, V. Mangano$^{9}$, L.A. Antonelli$^{8,10}$, P.
D'Avanzo$^{3,11}$, K. Hurley$^{12} $, I. F. Mirabel$^{13}$, L. J.
Pellizza$^{14}$, S. Piranomonte$^{8}$ \& L. Stella$^{8}$

\bigskip

\noindent$^{1}$INAF, Osservatorio Astrofisico di Arcetri, largo E. Fermi 5, I-50125 Firenze, Italy. \\
$^{2}$Dipartimento di Fisica, Universit\`a degli Studi di Milano-Bicocca, piazza delle Scienze 3, I-20126 Milano, Italy. \\
$^{3}$INAF, Osservatorio Astronomico di Brera, via E. Bianchi 46, I-23807 Merate (Lc), Italy. \\
$^{4}$Space Telescope Science Institute, 3700 San Martin Drive, Baltimore, Maryland 21218 Baltimore, USA. \\
$^{5}$Istituto Nazionale di Astrofisica, viale del Parco Mellini 84, I-00136 Roma, Italy. \\
$^{6}$Supernova Ltd., Olde Yord Village \#131, Northsound Road, Virgin Gorda, British Virgin Islands. \\
$^{7}$International School for Advanced Studies (SISSA/ISAS), via Beirut 2-4, I-34014 Trieste, Italy. \\
$^{8}$INAF, Osservatorio Astronomico di Roma, via di Frascati 33, I-00040 Monteporzio Catone (Roma), Italy. \\
$^{9}$INAF, Istituto di Astrofisica Spaziale e Fisica Cosmica di Palermo, via U. La Malfa 153, I-90146 Palermo, Italy. \\
$^{10}$ASI Science Data Center, via G. Galilei, I-00044 Frascati, Italy. \\
$^{11}$Dipartimento di Fisica e Matematica, Universit\`a dell'Insubria, via Valleggio 11, I-22100 Como, Italy. \\
$^{12}$University of California, Berkeley, Space Sciences Laboratory, Berkeley, California 94720-7450, USA. \\
$^{13}$European Southern Observatory, Alonso de C\'ordova 3107, Vitacura, Casilla 19001, Santiago 19, Chile. \\
$^{14}$Instituto de Astronom\'{\i}a y F\'{\i}sica del Espacio (CONICET/UBA), Casilla de Correos 67 Suc. 28, (1428) Buenos Aires, Argentina. \\

\bigskip

\noindent {\bf Gamma-ray bursts (GRBs) are short, intense flashes of soft
$\gamma$-rays coming from the distant Universe. Long-duration GRBs (those
lasting more than $\sim 2$~s) are believed to originate from the deaths of
massive stars$^1$, mainly on the basis of a handful of solid associations
between GRBs and supernovae$^{2-7}$. GRB\,060614, one of the closest GRBs
discovered, consisted of a 5-s hard spike followed by softer, brighter emission
that lasted for $\sim \mbox{}$100~s (refs 8, 9). Here we report deep optical
observations of GRB\,060614 showing no emerging supernova with absolute visual
magnitude brighter than $M_{\rm V} = -13.7$. Any supernova associated with
GRB\,060614 was therefore at least 100 times fainter, at optical wavelengths,
than the other supernovae associated with GRBs$^{10}$. This demonstrates that
some long-lasting GRBs can either be associated with a very faint supernova or
produced by different phenomena.} \bigskip

Following the discovery of the X-ray and optical afterglow of GRB\,060614$^9$,
we observed it with the European Southern Observatory (ESO) 8.2 m Very Large
Telescope (VLT). The spectrum of the host galaxy exhibits nebular emission
lines (Fig.~1), which reveal ongoing star formation and the presence of young,
massive stars. The specific star formation rate, normalized to the host
luminosity (B-band magnitude $M_{\rm B} = -15.5$), is about
$2~M_\odot$~yr$^{-1}~L_*^{-1}$, where $M_\odot$ is the solar mass and $L_*$ is
the typical luminosity of field galaxies. This value is comparable to that
exhibited by the Milky Way, and is at the low end of the distribution found for
long-duration GRB hosts$^{11}$. The galaxy is also fainter than most GRB
hosts$^{12}$.

Starting 15 h after the burst, we monitored the light curve of the optical
transient associated with GRB\,060614. Deep observations obtained in the R band
(roughly corresponding to the V band in the GRB rest frame) up to 65 d after
the burst did not reveal the emergence of a supernova component (Fig.~2; see
also refs 13, 14); such a component has been often observed in a number of
nearby long-duration GRBs. Our data are consistent with no supernova
contribution (blue line). Adopting as a template the light curve of
SN\,1998bw$^2$, we constrain the brightness of a supernova coincident with
GRB\,060614 to be at least 5.6 mag fainter than the template (3$\sigma$ limit;
green lines in Fig.~2). This corresponds to a peak absolute magnitude $M_{\rm
V} > -13.5$. Similar limiting magnitudes are obtained adopting different
supernova light curve shapes. A brighter supernova (yellow lines in Fig.~2)
would provide a totally inadequate fit. The faintness of a possible supernova
at optical wavelengths was further confirmed by a series of about ten spectra
obtained at the VLT in the 4,500--8,000~\AA{} wavelenght range, between 2006
June 15 and 2006 July 30. None of them shows the broad undulations due to the
very high expansion velocities ($\sim 30,000$ km~s$^{-1}$) typical of the
supernovae$^{15}$ associated with GRBs.

Such faintness cannot be due to dust extinction. First, the afterglow optical
spectra are not particularly red (from our BVRIJK photometry we measure a
spectral index $\beta = 0.94$ at $\sim 1.7$~d after the GRB). The afterglow is
also bright in the ultraviolet$^9$, where extinction would be more severe.
X-ray spectra (V.M. \textit{et al.}, manuscript in preparation) also show
little absorbing material along the line of sight (yielding a rest-frame
hydrogen column density $N_{\rm H} < 2 \times 10^{20}$ cm$^{-2}$ at 90\%
confidence level). Furthermore, by modelling the broad-band optical/X-ray
spectral energy distribution (see Supplementary Fig.~1), we can estimate that
the source was affected by less than 0.2 mag of extinction in the observed R
band. Allowing for this amount of extinction, we can refine our limit for the
supernova peak magnitude to $M_{\rm V} > -13.7$.

So far only type-Ib/c events have been clearly associated with GRBs$^1$ and,
had the progenitor of GRB\,060614 been one of them, its expected colour at
maximum light would be $B-V \approx 0.5 \pm 0.1$, as measured in well-observed
events such as SN\,2006aj$^{10}$, SN\,2002ap$^{16}$ and SN\,1998bw$^{2}$, as
well in several other type-Ib/c SNe. From the tables of ref.~17, we find that
this colour corresponds to an effective temperature $T \approx 6,500$~K, which,
combined with the lower limit to the absolute magnitude, provides a strict
upper limit to the bolometric luminosity, at maximum light, of $L < 10^{41}$
erg~s$^{-1}$. Therefore the radius of the emitting region is constrained to be
$R \approx \sqrt{L/(4\pi \sigma T^4)} < 2.8 \times 10^{14}$~cm (where $\sigma$
is the Stefan-Boltzmann constant). As the rise times to maximum light of
type-Ib/c supernovae range between 10 and 20~d (refs 10, 18-20) in the V band,
the upper limit to the radius implies an upper limit to the expansion velocity
in the range 1,600--3,200 km~s$^{-1}$. This value is an order of magnitude
smaller than that observed in SNe associated with GRBs$^{7,15}$.

The low expansion velocity and the faint luminosity implied for a possible
supernova progenitor are reminiscent of a class of very faint core-collapse
supernovae recently discovered$^{21}$ in the local Universe. They are of type
II, have absolute magnitudes at maximum in the range $-13 > M_{\rm V} > -15$,
and show very small expansion velocities ($\sim 1,000$ km~s$^{-1}$). The
properties of such objects may well be consistent with the available data on
GRB\,060614.

Faint type-II SNe have been interpreted in terms of the collapse of massive
stars with an explosion energy so small that most of $^{56}$Ni falls back onto
the compact stellar remnant$^{22}$. Such supernovae share properties with the
present case, both in terms of observational characteristics and because they
are expected to give rise to a black hole (which is believed to be necessary
for the production of a GRB). However, the possibility that such supernova
progenitors are able to produce GRBs has yet to be explored. In particular, the
stellar envelope would need to be absent for the relativistic jets to emerge
out of the star. GRB\,060614 might thus be an example of a fallback supernova
of type Ib/c. The small amount of nickel ($< 10^{-3}\, M_\odot$), responsible
for the very low luminosity, might possibly also provide little heating to the
ejecta, leading to an unusually low temperature ($T \approx 2,000$~K) and
allowing for larger velocities (as $v \propto T^{-2}$). In any case,
GRB\,060614 may be the prototype of a new class of GRBs originating from a new
kind of massive stars death, different from those producing both classical
long-duration (associated with bright type-Ib/c supernovae) and short-duration
(possibly originating in binary system mergers$^{23}$) GRBs. Some evidence for
this idea comes from the high-energy properties of this GRB, which
contemporaneously exhibits features typical of both the long and short GRB
classes$^8$. Indeed, scenarios in which the GRB was not directly connected to a
supernova explosion$^{24}$ cannot be excluded by our data (though they are not
required). For example, our data would be compatible with a supernova exploding
before the GRB$^{25}$. Also, a binary merger mechanism$^{26}$, similar to that
proposed to power short-duration GRBs, or some type of collapsar model$^{27}$,
are not expected to produce a supernova. These results challenge the commonly
accepted scenario, in which long-duration GRBs are produced only together with
very bright supernova explosions. Not all GRBs are produced in such a way.

\bigskip

\begin{enumerate}

\item Woosley, S. E. \& Bloom, J. S. The supernova-gamma-ray burst connection. \textit{Annu. Rev. Astron. Astrophys.} \textbf{44}, 507-556 (2006).

\item Galama, T. J. \textit{et al.} An unusual supernova in the error box of the $\gamma$-ray burst of 25 April 1998. \textit{Nature} \textbf{395}, 670-672 (1998).

\item Stanek, K. Z. \textit{et al.} Spectroscopic discovery of the supernova 2003dh associated with GRB 030329. \textit{Astrophys. J.} \textbf{591}, L17-L20 (2003).

\item Hjorth, J. \textit{et al.} A very energetic supernova associated with the $\gamma$-ray burst of 29 March 2003. \textit{Nature} \textbf{423}, 847-850 (2003).

\item Malesani, D. \textit{et al.} SN 2003lw and GRB 031203: A bright supernova for a faint gamma-ray burst. \textit{Astrophys. J.} \textbf{609}, L5-L8 (2004).

\item Pian, E. \textit{et al.} An optical supernova associated with the X-ray flash XRF 060218. \textit{Nature} \textbf{442}, 1011-1013 (2006).

\item Campana, S. \textit{et al.} The association of GRB 060218 with a supernova and the evolution of the shock wave. \textit{Nature} \textbf{442}, 1008-1010 (2006).

\item Gehrels, N. \textit{et al.} A new $\gamma$-ray burst classification scheme from GRB\,060614. \textit{Nature}, \textbf{444}, 1044-1046 (2006).

\item Parsons, A. M. \textit{et al.} GRB 060614: Swift detection of a burst with a bright optical and X-ray counterpart. \textit{GCN Circ.} \textbf{5252} (2006).

\item Ferrero, P. \textit{et al.} The GRB060218/SN 2006aj event in the context of other gamma-ray burst supernovae. \textit{Astron. Astrophys.} \textbf{457}, 857-864 (2006).

\item Christensen, L., Hjorth, J. \& Gorosabel, J. UV star-formation rates of GRB host galaxies. \textit{Astron. Astrophis.} \textbf{425}, 913-926 (2004).

\item Fruchter, A. S. \textit{et al.} Long $\gamma$-ray bursts and core-collapse supernovae have different environments. \textit{Nature}, \textbf{441}, 463-468 (2006).

\item Gal-Yam, A. \textit{et al.} A novel explosive process is required for the $\gamma$-ray burst GRB\,060614. \textit{Nature}, \textbf{444}, 1053-1055 (2006).

\item Fynbo, J. P. U. \textit{et al.} No supernovae associated with two long-duration $\gamma$-ray bursts. \textit{Nature}, \textbf{444} 1047-1049 (2006).

\item Patat, F. \textit{et al.} The metamorphosis of SN 1998bw. \textit{Astrophys. J.}, \textbf{555}, 900-917 (2001).

\item Foley, R. \textit{et al.} Optical photometry and spectroscopy of the SN 1998bw-like type Ic supernova 2002ap. \textit{Publ. Astron. Soc. Pacif.} \textbf{115}, 1220-1235 (2003).

\item Romaniello, M., Panagia, N., Scuderi, S. \& Kirshner, R. P. Accurate stellar population studies from multiband photometric observations. \textit{Astronom. J.} \textbf{123}, 915-940 (2002).

\item Della Valle, M. \textit{et al.} Hypernova signatures in the late rebrightening of GRB 050525A. \textit{Astrophys. J.} \textbf{642}, L103-L106 (2006).

\item Hamuy, M. Observed and physical properties of core-collapse supernovae. \textit{Astrophys. J.} \textbf{582}, 905-914 (2003).

\item Panagia, N. in \textit{Supernovae and Gamma-Ray Bursters} (ed. Weiler, K. W.) 113-144 (Lecture Notes in Physics, Vol. 598, Springer, Berlin, 2003). 

\item Pastorello A. \textit{et al.} Low luminosity type II supernovae: Spectroscopic and photometric evolution. \textit{Mon. Not. R. Astron. Soc.} \textbf{347}, 74-94 (2004).

\item Nomoto, K. \textit{et al.} in \textit{Stellar Collapse} (ed. Fryer, C.) 277-325 (Astrophysics and Space Science Library, Vol. 302, Kluwer Academic, Dordrecht, 2004).

\item Eichler, D., Livio, M., Piran, T. \& Schramm, D. N. Nucleosynthesis, neutrino bursts and $\gamma$-rays from coalescing neutron stars. \textit{Nature}, \textbf{340}, 126-128 (1989).

\item Blinnikov, S.I. \& Postnov, K.A. A mini-SN model for optical afterglow of GRBs. \textit{Mon. Not. R. Astron. Soc.} \textbf{293}, L29-L32 (1998).

\item Vietri, M. \& Stella, L. A gamma-ray burst model with small baryon contamination. \textit{Astrophis. J.} \textbf{507}, L45-L48 (1998).

\item Belczynski, K., Bulik, T. \& Rudak, B. Study of gamma-ray burst binary progenitors. \textit{Astrophys. J.} \textbf{571}, 394-412 (2002).

\item Woosley, S.E., Zhang, W. \& Heger, A. in \textit{Gamma-Ray Burst and Afterglow Astronomy 2001: A Workshop Celebrating the First Year of the HETE Mission} (eds Ricker, G. R. \& Vanderspek, R.) 185-192 (AIP Conf. Proc., Vol. 662, American Institute of Physics, New York, 2003).

\item Price, P. A., Berger, E. \& Fox, D. B. GRB 060614: Redshift. \textit{GCN Circ.} \textbf{5275} (2006).

\item French, F., Melady, G., Hanlon, L., Jel\'\i{}nek, M. \& Kub\'anek, P. GRB060614: Watcher observation \textit{GCN Circ.} \textbf{5257} (2006).

\end{enumerate}

\bigskip

\noindent \textbf{Supplementary information} is linked to the online version of
the paper at \texttt{www.nature.com/nature}.\bigskip

\noindent \textbf{Acknowledgments.} This work is based on data collected at the
Very Large Telescope operated by the European Southern Observatory. We
acknowledge support from the observing staff. This research is supported in
Italy by ASI, MIUR-PRIN and INAF-PRIN grants.\bigskip

\noindent \textbf{Author information.} Reprint and permissions information is
available at \texttt{www.nature.com/reprints}. The authors declare no competing
financial interests. Correspondence and requests for materials should be
addressed to M.D.V. (\texttt{massimo@arcetri.astro.it}).\bigskip

\clearpage

\begin{center}
  \includegraphics[width=0.8\columnwidth]{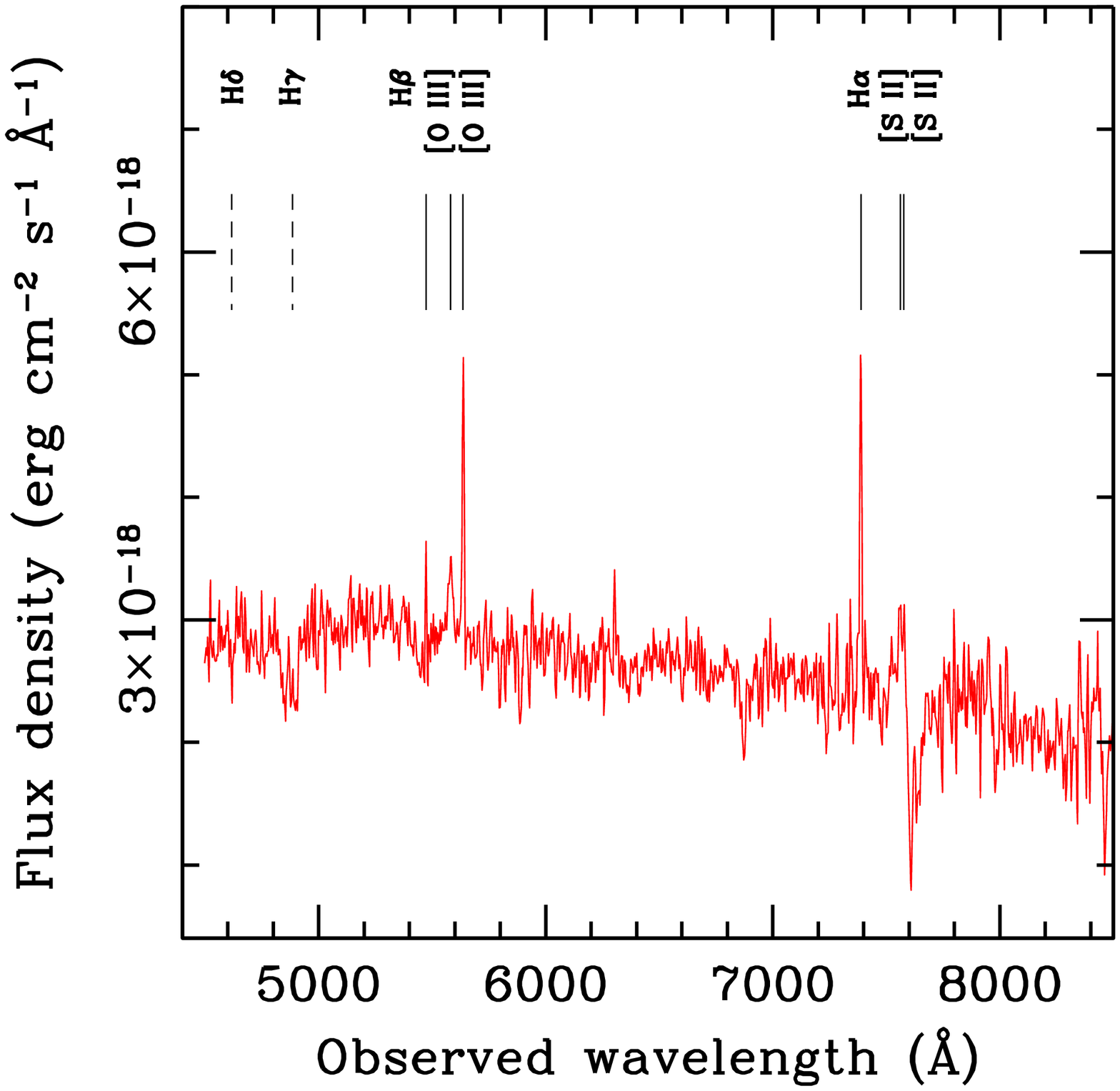}
\end{center}

\noindent {\bf Figure 1 $|$ Spectrum of the host galaxy of GRB\,060614.} This
is the average of several observations taken with the VLT equipped with the
FORS\,2 spectrograph in the period 2006 June 20 to July 30. From the emission
features (marked with solid bars) we infer $z = 0.1254 \pm 0.0005$. This
confirms the redshift proposed in ref.~28. H$\gamma$ and H$\delta$ are seen in
absorption (dashed bars). The flux from the H$\alpha$ line, not corrected for
internal extinction, amounts to $4.1 \times 10^{-17}$ erg~cm$^{-2}$~s$^{-1}$
(corrected for slit loss). This corresponds to an unobscured star formation
rate of $1.3 \times 10^{-2}~M_\odot$~yr$^{-1}$. Given the faintness of the
galaxy ($M_{\rm B} \approx -15.5$), however, the specific star formation rate
($2\, M_\odot$~yr$^{-1}~L_*^{-1}$, assuming for the absolute B-band magnitude
of field galaxies $M^*_B = -21$), is not negligible. From the observed flux of
the [O\,III] lines and the limits on [N\,II], we infer a metallicity larger
than $\sim 1/20$ solar.

\includegraphics[width=\columnwidth]{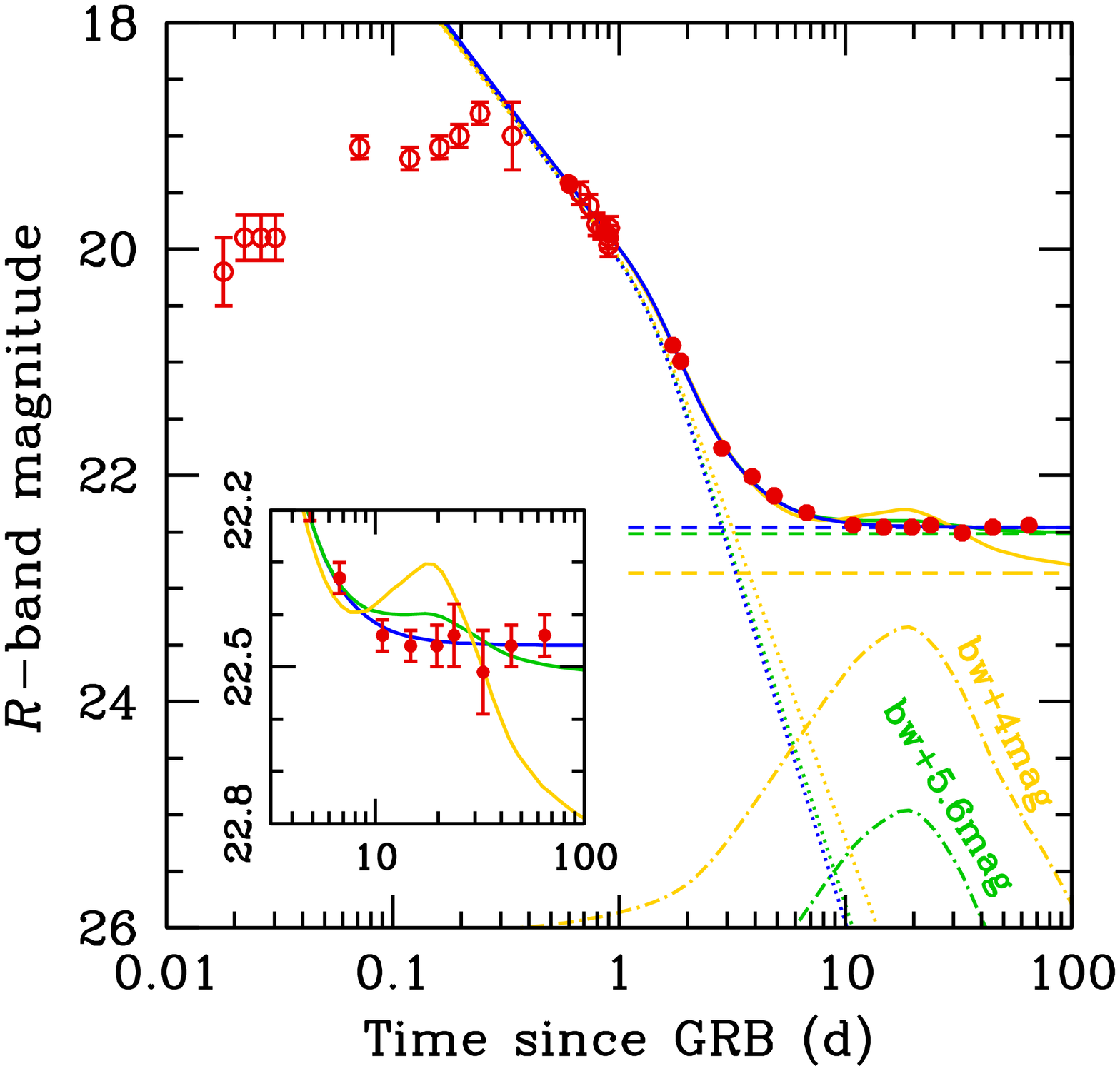}

\noindent {\bf Figure 2 $|$ R-band light curve of the GRB\,060614 afterglow.}
Red open circles show data from the literature$^{13,14,29}$ (not used in the
fits); red filled circles represent our VLT data (see Supplementary Table 1).
Error bars (smaller than symbols for most of our data) show the 1$\sigma$
errors. Photometry was performed adopting large apertures in order to include
all the flux from the host galaxy. Flux calibration was achieved by observing
several Landolt standard fields. The data were modelled as the sum (solid
lines) of three components: the afterglow (dotted lines), the host (dashed
lines) and a supernova akin to SN\,1998bw but rescaled in flux (`bw';
dot-dashed lines). The different colours correspond to different contributions
from the supernova: no contribution (blue), a supernova fainter by 5.6 mag
(green), and a supernova fainter by 4 mag (yellow). The model shown by gren
lines corresponds to the brightest supernova allowed by our data, $M_{\rm V} >
-13.5$, at the 3$\sigma$ level. At the 2$\sigma$ level, the limit is $M_{\rm V}
> -12.9$. The model shown by the yellow lines is clearly inadequate. The inset
shows an expanded version of the plot around the peak of a putative supernova.
The afterglow component is described by a broken power law with decay indices
$\alpha_1 = 1.08 \pm 0.03$ and $\alpha_2 = 2.48 \pm 0.07$, respectively before
and after the break at $t_{\rm break} = 1.39 \pm 0.04$ d ($\chi^2/{\rm d.o.f} =
15.5/20$). Interpreting this as a jet break, the inferred jet half-opening
angle is $\vartheta \approx 12^\circ$.

\clearpage

\includegraphics[width=\columnwidth]{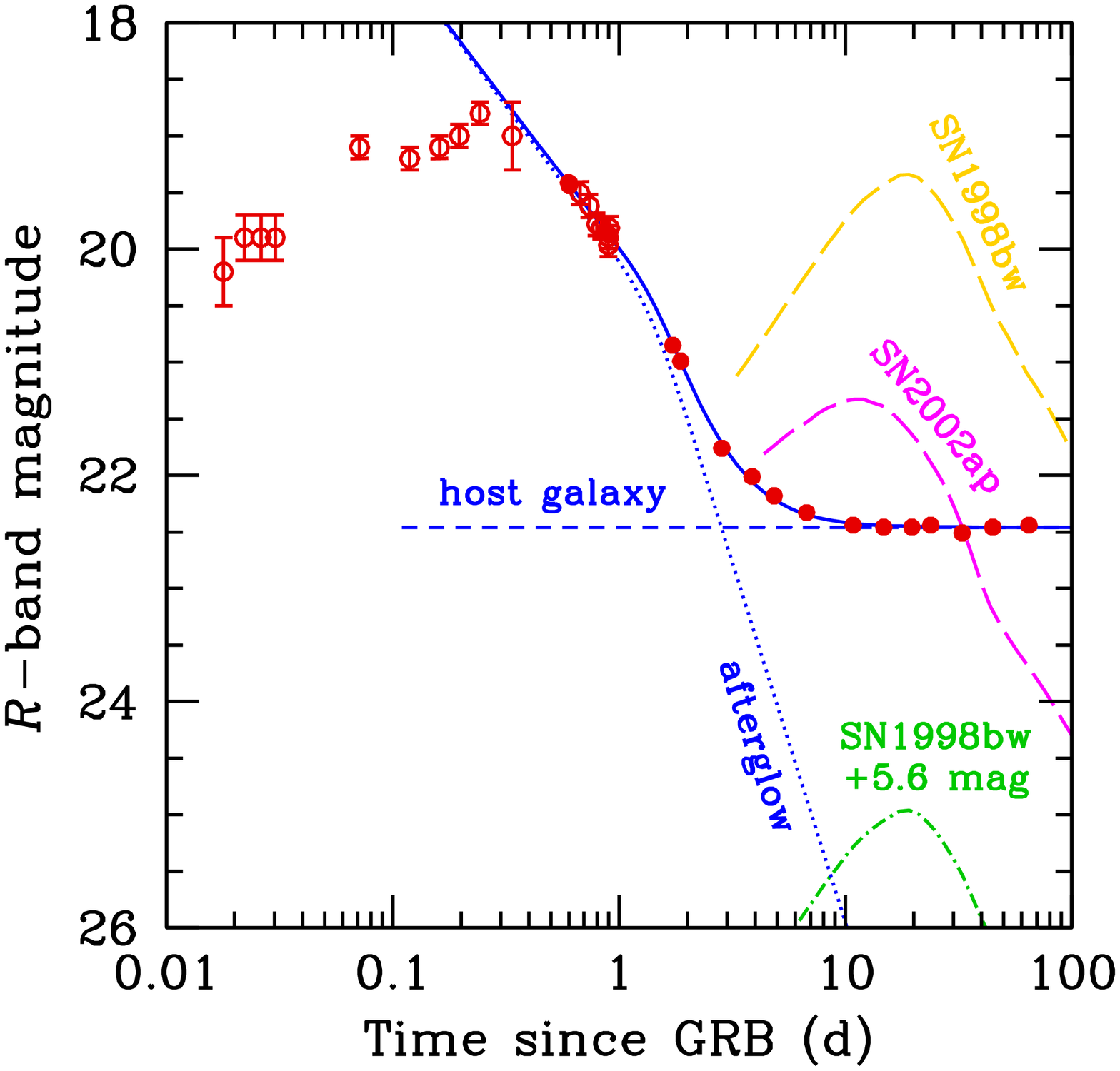}

\noindent {\bf Extra figure $|$ Comparison of the light curve of the
GRB\,060614 afterglow and those of SN\,1998bw and SN\,2006aj}. The yellow and
magenta long-dashed lines show the light curves of two supernovae previously
found to be associated with other GRBs reported at $z = 0.1254$. The green
dot-dashed line shows the brightest supernova allowed by our data at the
3$\sigma$ level. The contribution of the host galaxy (short-dashed line) and of
the afterglow (dotted line) are also indicated. The data are the same as in
Fig.~2.

\clearpage

\begin{center}
  \bf\Large Supplementary material for ``An enigmatic long-lasting $\gamma$-ray
  burst not accompanied by a bright supernova''.
\end{center}

\bigskip

\includegraphics[width=\textwidth]{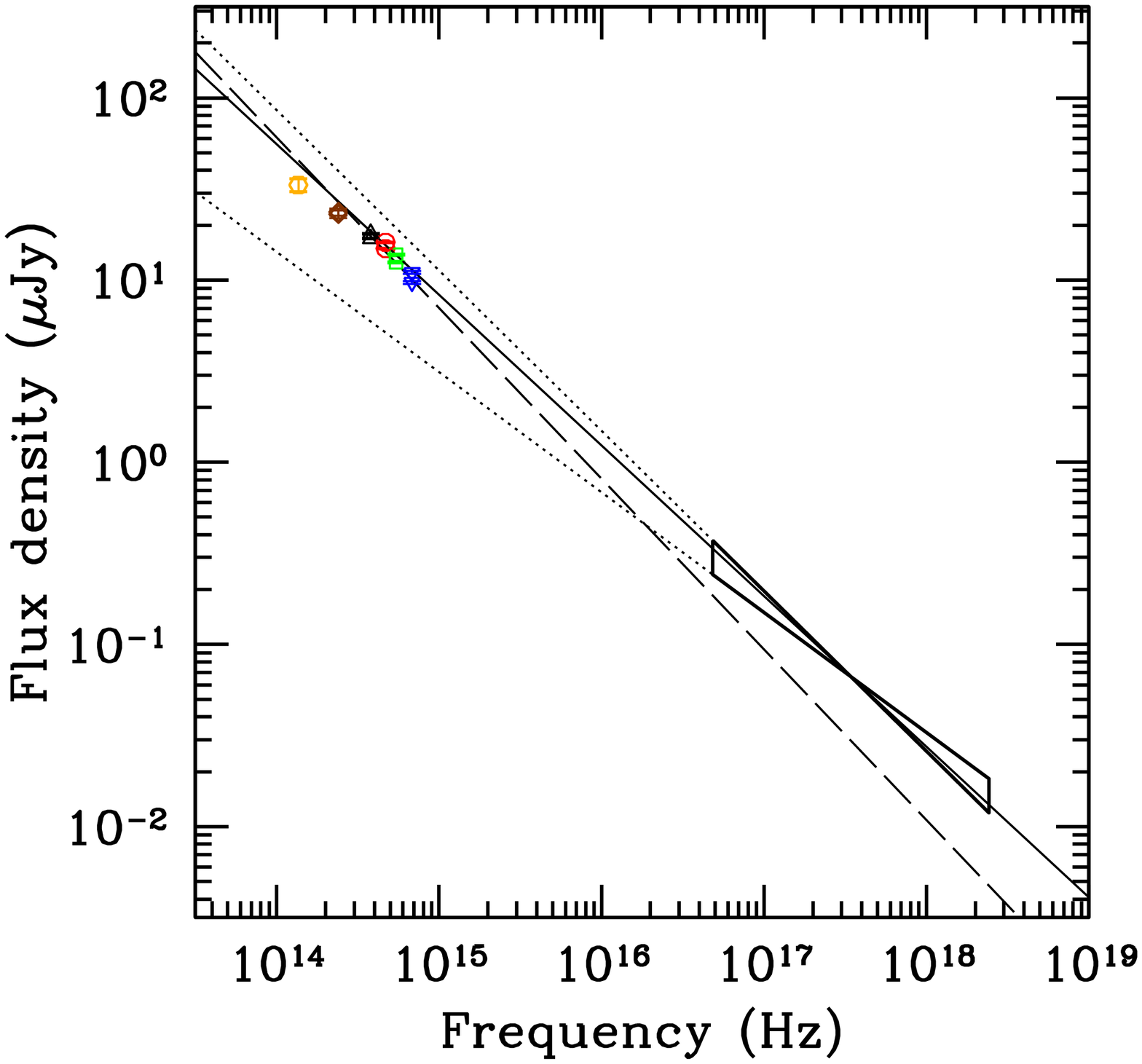}

\noindent {\bf Supplementary Figure 1 $|$ Estimate of the reddening
correction.} We interpret the afterglow radiation as due to synchrotron
emission, as usually assumed for GRB afterglows. In order to build the optical
spectral energy distribution (SED), we adopt our \textit{BVRIJK} photometry and
take the X-ray data from the work by V.M. \textit{et al.} (in preparation).
Using our best-fit light curve we report all data at the common epoch $t =
1.73$ days after the burst (the time around which our measurements cluster).
After correcting for the Galactic extinction ($A_V = 0.07$ mag), the optical
spectrum is well fit by a power law ($F_\nu \propto \nu^{-\beta}$) with index
$\beta = 0.94 \pm 0.08$ (dashed line). At the same time, the X-ray spectrum is
described by a power law with index $\beta = 0.77 \pm 0.11$. Supplementary
Figure 1 shows the broad-band SED at this epoch. As can be seen, the optical
and X-ray data belong to the same power-law segment. The common origin of the
two components is also confirmed by their similar temporal properties. Assuming
an SMC-like extinction curve, to make the optical data match exactly the X-ray
flux, an amount of extinction $A_V = 0.08$ mag is required. Since the
rest-frame $V$ band corresponds to the $R$ band, this is also the amount of
extinction suffered by our measurements. Assuming different extinction curves,
the absorption corresponding to the observed $R$ band does not change
significantly and is always less than 0.1 mag.

\newpage

\noindent {\bf Supplementary Table 1 $|$ Log of observations}. The time $t_0$
indicates the GRB trigger time, 2006 June 14.53042 UT. Errors are given at the
1-$\sigma$ confidence level.\bigskip

{\scriptsize 

\noindent\begin{tabular}{llllllll} \hline \hline
Mean date         & $t-t_0$    & Filter & Exposure      & Airmass & Seeing & Instrument & Magnitude      \\
(UT)              & (days)     &	& (s)		&	  & ($''$) &		&		 \\ \hline \hline
2006 Jun 16.27492 &  1.74450   & $B$	& 4$\times$90	& 1.23    & 0.60   & VLT+FORS1  & 21.72$\pm$0.02 \\
2006 Sep 02.05099 & 79.52057   & $B$	& 10$\times$120 & 1.25    & 1.30   & VLT+FORS1  & 23.62$\pm$0.15 \\ \hline
2006 Jun 16.25076 &  1.72034   & $V$	& 2$\times$120  & 1.30    & 0.55   & VLT+FORS1  & 21.20$\pm$0.01 \\
2006 Jun 17.36557 &  2.83515   & $V$	& 4$\times$90	& 1.14    & 0.50   & VLT+FORS1  & 22.16$\pm$0.02 \\
2006 Jun 18.39119 &  3.86077   & $V$	& 6$\times$180  & 1.17    & 0.80   & VLT+FORS1  & 22.47$\pm$0.02 \\
2006 Jul 08.33006 & 23.79964   & $V$	& 2$\times$120  & 1.16    & 1.20   & VLT+FORS1  & 22.75$\pm$0.07 \\
2006 Jul 17.31686 & 32.78644   & $V$	& 5$\times$120  & 1.17    & 1.10   & VLT+FORS1  & 22.74$\pm$0.09 \\ \hline
2006 Jun 15.12757 &  0.59715   & $R$	& 1$\times$20	& 1.16    & 2.44   & VLT+FORS2  & 19.41$\pm$0.03 \\
2006 Jun 15.12927 &  0.59885   & $R$	& 1$\times$60	& 1.22    & 2.41   & VLT+FORS2  & 19.42$\pm$0.02 \\
2006 Jun 15.13031 &  0.59989   & $R$	& 1$\times$60	& 1.22    & 2.39   & VLT+FORS2  & 19.41$\pm$0.02 \\
2006 Jun 15.13136 &  0.60094   & $R$	& 1$\times$60	& 2.44    & 2.37   & VLT+FORS2  & 19.43$\pm$0.02 \\
2006 Jun 15.13242 &  0.60200   & $R$	& 1$\times$60	& 2.41    & 2.35   & VLT+FORS2  & 19.43$\pm$0.02 \\
2006 Jun 15.13355 &  0.60313   & $R$	& 1$\times$60	& 2.39    & 2.33   & VLT+FORS2  & 19.42$\pm$0.02 \\
2006 Jun 15.13460 &  0.60418   & $R$	& 1$\times$60	& 2.37    & 2.31   & VLT+FORS2  & 19.44$\pm$0.02 \\
2006 Jun 15.13566 &  0.60524   & $R$	& 1$\times$60	& 2.35    & 2.29   & VLT+FORS2  & 19.43$\pm$0.02 \\
2006 Jun 15.13672 &  0.60630   & $R$	& 1$\times$60	& 2.33    & 2.27   & VLT+FORS2  & 19.43$\pm$0.02 \\
2006 Jun 15.40052 &  0.87010   & $R$	& 1$\times$10	& 2.31    & 1.16   & VLT+FORS2  & 19.83$\pm$0.03 \\
2006 Jun 15.43038 &  0.89996   & $R$	& 1$\times$60	& 2.27    & 1.22   & VLT+FORS2  & 19.87$\pm$0.02 \\
2006 Jun 16.25625 &  1.72583   & $R$	& 2$\times$120  & 1.28    & 1.22   & VLT+FORS1  & 20.85$\pm$0.01 \\
2006 Jun 16.40016 &  1.86974   & $R$	& 2$\times$120  & 1.17    & 0.45   & VLT+FORS1  & 20.99$\pm$0.01 \\
2006 Jun 17.37241 &  2.84199   & $R$	& 2$\times$120  & 1.14    & 0.50   & VLT+FORS1  & 21.76$\pm$0.02 \\
2006 Jun 18.39941 &  3.86899   & $R$	& 6$\times$180  & 1.19    & 0.65   & VLT+FORS1  & 22.01$\pm$0.02 \\
2006 Jun 19.37407 &  4.84365   & $R$	& 2$\times$180  & 1.15    & 0.50   & VLT+FORS1  & 22.18$\pm$0.04 \\
2006 Jun 21.27125 &  6.74083   & $R$	& 3$\times$180  & 1.20    & 0.60   & VLT+FORS1  & 22.33$\pm$0.03 \\
2006 Jun 25.34483 & 10.81441   & $R$	& 2$\times$300  & 1.14    & 0.90   & VLT+FORS1  & 22.44$\pm$0.03 \\
2006 Jun 29.30301 & 14.77259   & $R$	& 11$\times$180 & 1.16    & 0.50   & VLT+FORS1  & 22.46$\pm$0.03 \\
2006 Jul 04.20860 & 19.67818   & $R$	& 6$\times$240  & 1.28    & 0.60   & VLT+FORS1  & 22.46$\pm$0.04 \\
2006 Jul 08.33536 & 23.80494   & $R$	& 2$\times$120  & 1.16    & 1.40   & VLT+FORS1  & 22.44$\pm$0.06 \\
2006 Jul 17.32709 & 32.79667   & $R$	& 3$\times$180  & 1.19    & 1.10   & VLT+FORS1  & 22.51$\pm$0.08 \\
2006 Jul 29.26643 & 44.73601   & $R$	& 5$\times$240  & 1.15    & 0.90   & VLT+FORS1  & 22.46$\pm$0.04 \\
2006 Aug 18.23409 & 64.70367   & $R$	& 12$\times$300 & 1.18    & 0.60   & VLT+FORS1  & 22.44$\pm$0.04 \\ \hline
2006 Jun 16.26278 &  1.73236   & $I$	& 3$\times$120  & 1.26    & 0.55   & VLT+FORS1  & 20.49$\pm$0.01 \\
2006 Jun 17.37868 &  2.84826   & $I$	& 3$\times$120  & 1.15    & 0.43   & VLT+FORS1  & 21.25$\pm$0.02 \\
2006 Jun 18.38882 &  3.85840   & $I$	& 4$\times$300  & 1.16    & 0.75   & VLT+FORS1  & 21.50$\pm$0.02 \\
2006 Jul 08.34050 & 23.81008   & $I$	& 2$\times$120  & 1.17    & 1.10   & VLT+FORS1  & 21.70$\pm$0.06 \\
2006 Jul 17.33614 & 32.80572   & $I$	& 3$\times$180  & 1.20    & 0.95   & VLT+FORS1  & 21.76$\pm$0.07 \\ \hline
2006 Jun 15.30022 &  0.76980   & $J$	& 15$\times$60  & 1.13    & 1.20   & NTT+SofI	& 18.51$\pm$0.02 \\ \hline
2006 Jun 15.28454 &  0.75411   & $K$	& 8$\times$60	& 1.16    & 1.15   & NTT+SofI	& 17.23$\pm$0.07 \\
2006 Jun 15.29074 &  0.76032   & $K$	& 7$\times$60	& 1.15    & 1.15   & NTT+SofI	& 17.26$\pm$0.08 \\
2006 Jun 16.20839 &  1.67797   & $K$	& 10$\times$60  & 1.40    & 1.10   & NTT+SofI	& 18.18$\pm$0.09 \\ \hline \hline
\end{tabular}

}

\end{document}